\pdfoutput=1
\RequirePackage{pgfplotstable}
\documentclass[sn-basic]{sn-jnl}
\jyear{2023}%
\theoremstyle{thmstyleone}%
\theoremstyle{thmstyletwo}%

\theoremstyle{thmstylethree}%

\usepackage{amsmath}
\usepackage{amsfonts}

\usepackage{graphicx}
\graphicspath{ {figures/} }

\usepackage{booktabs}
\usepackage{tabularx}
\usepackage{multirow}

\usepackage{array}
\newcolumntype{Y}[1]{>{\centering\arraybackslash}X}
\usepackage{makecell}
\usepackage{tikz}
\usetikzlibrary{arrows,positioning} 

\usepackage{xurl}

\begin{document}

\title[Online Privacy Disclosure Detection with Multi-Label Text Classification]{When Graph Convolution Meets Double Attention: Online Privacy Disclosure Detection with Multi-Label Text Classification}

\author*[1]{\fnm{Zhanbo} \sur{Liang}}\email{doctorliang@sjtu.edu.cn}
\author*[1]{\fnm{Jie} \sur{Guo}}\email{guojie@sjtu.edu.cn}
\author[1]{\fnm{Weidong} \sur{Qiu}}\email{qiuwd@sjtu.edu.cn}
\author[1]{\fnm{Zheng} \sur{Huang}}\email{huang-zheng@sjtu.edu.cn}
\author*[2]{\fnm{Shujun} \sur{Li}}\email{s.j.li@kent.ac.uk}

\affil[1]{\orgdiv{School of Cyber Science and Engineering}, \orgname{Shanghai Jiao Tong University}, \orgaddress{\street{800 Dongchuan Road, Minhang District}, \city{Shanghai}, \postcode{200240}, \country{China}}}

\affil[2]{\orgdiv{Institute of Cyber Security for Society (iCSS) \& School of Computing}, \orgname{University of Kent}, \orgaddress{\city{Canterbury}, \postcode{CT2 7NP}, \state{Kent}, \country{UK}}}

\abstract{With the rise of Web 2.0 platforms such as online social media, people's private information, such as their location, occupation and even family information, is often inadvertently disclosed through online discussions. Therefore, it is important to detect such unwanted privacy disclosures to help alert people affected and the online platform. In this paper, privacy disclosure detection is modeled as a multi-label text classification (MLTC) problem, and a new privacy disclosure detection model is proposed to construct an MLTC classifier for detecting online privacy disclosures. This classifier takes an online post as the input and outputs multiple labels, each reflecting a possible privacy disclosure. The proposed presentation method combines three different sources of information, the input text itself, the label-to-text correlation and the label-to-label correlation. A double-attention mechanism is used to combine the first two sources of information, and a graph convolutional network (GCN) is employed to extract the third source of information that is then used to help fuse features extracted from the first two sources of information. Our extensive experimental results, obtained on a public dataset of privacy-disclosing posts on Twitter, demonstrated that our proposed privacy disclosure detection method significantly and consistently outperformed other state-of-the-art methods in terms of all key performance indicators.}

\keywords{Privacy disclosure detection, User generated content (UGC), Online social media, Graph convolutional network (GCN), Multi-label text classification (MLTC)}

\maketitle

\section{Introduction}

\begin{figure}[!htb]
\centering
\includegraphics[width=0.8\linewidth]{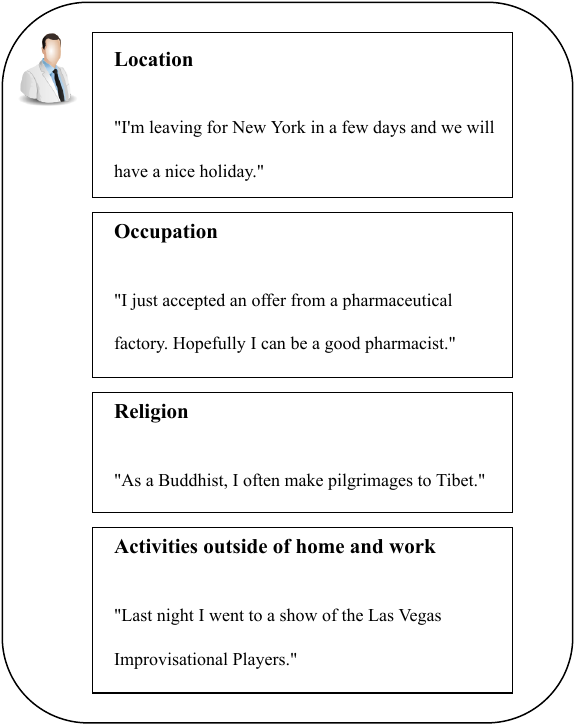}
\caption{Several illustrative examples of possible privacy disclosures on online social media platforms.}
\label{fig:twitter_case}
\end{figure}

The rapid development of information and communication technologies have helped facilitate people's social interactions. Online social media platforms like Twitter provide people a new way to build up their social relationships, share their daily lives, and express their emotions. However, many online users frequently (and often unintentionally) share personal information online, which can lead to unwanted online disclosures of private information of themselves or other people in their social networks. Figure~\ref{fig:twitter_case} shows several imaginary but realistic online posts of such unintended privacy disclosures on Twitter, generated based on some examples in a research dataset of privacy-disclosing tweets constructed by~\citet{song2018personal}. Although people can check their online posts manually to avoid privacy disclosures, many online users do not have a good level of awareness on such privacy issues, and they do not necessarily know when and what to check. Therefore, automated solutions that can help online users identify such issues and take proper actions are important, which is the focus of our work.

Past studies about privacy disclosure detection attempted to solve this problem with different machine learning methods. Traditional methods on privacy disclosure detection try to detect privacy disclosures in user profiles or user settings, but not in user generated content (UGC), leading to incomplete detection. More recently, many researchers started studying privacy disclosure detection in UGC by analysing pictures and/or texts in such UGC. Therefore, their work extends the scope of such work.

Recently, some researchers use the multi-label text classification (MLTC) framework to model the privacy disclosure problem~\citep{song2018personal, Chen2020GrHA}. MLTC is a an important task in the field of natural language processing (NLP). Different from multi-class text classification (MCTC), which classifies a given piece of text into one of multiple class labels, MLTC aims to tag a piece of given text with \emph{multiple} (i.e., one or more) content-specific labels. In~\citep{song2018personal, Chen2020GrHA}, the privacy information is divided into eight main categories, then they make further division, using 32 categories of labels to reflect the possible disclosed privacy. However, their methods are limited due to the lack of consideration for the relationship between texts and labels. Their methods aim to improve the prediction results by considering the co-occurrence relation between labels. For example, the label ``Health condition" usually appears with the label ``Treatment" and the label ``Occupation" usually appears with the label ``Salary". However, those two methods do not consider label-text correlations, i.e., their work ignores the fact that some key words or phrases in the input texts can assist indicating the possible privacy-aware labels. For example, a location name in the input text may help to indicate that the text is involved in the privacy disclosure of ``Current location" or ``Place planning to go". We follow their thoughts to model privacy disclosure detection as an MLTC problem. Our proposed framework takes an online post as the input, and outputs a number of privacy-relevant labels that indicate potential disclosure of different types of personal information in the input online post.

Considering that privacy disclosure is a universal problem in people's daily life, new frameworks with better performance on privacy disclosure detection are needed. The aim of our work is to provide a more effective MLTC privacy disclosure detection algorithm to facilitate the fine-grained text privacy detection. As mentioned before, current MLTC privacy disclosure models are limited by their consideration of relationships between various texts or words. In order to improve the performance of privacy-disclosing post detection, which combines three different sources of relevant information, the text information, the label-to-text correlation and the label-to-label correlation, to produce a more comprehensive model for detecting privacy-disclosing online posts. Our model extracts the text representations through a double-attention mechanism as~\citet{LSAN_2019} did, which measures the contribution of each word to each privacy-relevant label. The label-to-label correlation is considered in the final text representation via a graph convolutional network (GCN). We propose a new feature fusion mechanism assisted by GCN to make the fused feature more comprehensive. We utilize the label-to-label correlation to obtain the proposed compensation coefficients from both the self-attention and the label-attention text representations. We summarize the main contributions of our work as follows:
\begin{itemize}
\item A new privacy disclosure detection model with multi-label text classification is proposed. Our model presents a new fine-grained privacy disclosure detection algorithm and outputs multiple privacy-aware labels as the possible leaked privacy. From the perspective of the detection performance, our model provides a better solution to the fine-grained privacy disclosure detection on the UGC.

\item Our proposed model considers three different sources of relevant information for the MLTC task: the input text itself, the label-to-text correlation, and the label-to-label correlation.

\item A new feature fusion mechanism assisted by a GCN is proposed to construct comprehensive text representations with the guidance of the label-to-label correlation. The idea of compensation coefficients is proposed in the feature fusion mechanism, which reflects the compensation relationship between self-attention and label-attention.

\item A series of experiments on a public privacy-disclosing tweet dataset showed that our proposed model outperformed selected state-of-the-art models significantly and consistently. Our code has been released to facilitate others to conduct follow-up research.\footnote{\url{https://github.com/xiztt/wgma}}
\end{itemize}

The rest of the paper is organized as follows. Section~\ref{sec:related_work} introduces the related work. Section~\ref{sec:proposed_method} elaborates the proposed MLTC-based model for privacy detection. Section~\ref{sec:experimental_results} shows and discusses the experiment results. Section~\ref{sec:conclusion} concludes our work and discusses the future work. Section~\ref{sec:statements} makes statements on financial or non-financial interests that are directly or indirectly related to the work submitted for publication.

\section{Related Work}
\label{sec:related_work}

\subsection{Privacy Disclosure Analysis}

The problem of online privacy disclosures has attracted the attention of many researchers. Some researchers studied this problem based on analysis of user profiles~\citep{2017Privacy, eslami2017privacy, huang2019neural} or privacy settings of user accounts~\citep{raber2018deriving, sanchez2020recommendation}. \citet{2017Privacy} proposed a privacy-aware framework that leverages solidarity in a large community to scramble user interaction histories, in order to disturb the information collection from user profiles by the online service providers. To minimize users' privacy risks, \citet{eslami2017privacy} proposed an alternative solution, where posts of different users are split and merged into synthetic mediator profiles. \citet{raber2018deriving} studied privacy settings of user accounts by observing the context factors and personality measures which can be used to predict the correct privacy level out of seven privacy levels. \citet{sanchez2020recommendation} considered how to model users' privacy preferences for data sharing and processing in the IoT and fitness domain, paying a specific attention to the GDPR compliance.

Some other researchers such as~\citet{tran2016privacy} and \citet{mao2011loose} also proposed classifiers to detect privacy disclosures in user-generated online posts. \citet{tran2016privacy} proposed Privacy-CNH, a binary classification framework that utilizes hierarchical features including both object and convolutional features in a deep learning model to detect whether a photo is private or not. \citet{mao2011loose} analysed privacy disclosures on Twitter by building binary classifiers to detect three types of privacy disclosure including divulging vacation plans, tweeting under the influence of alcohol and revealing medical conditions. Despite all the past studies, they only focused on privacy disclosure detection at a more coarse-grained level. These studies used frameworks or classifiers to implement relatively simple analysis of privacy disclosures, normally based on less comprehensive privacy categories so not being able to cover some specific privacy disclosure scenarios.

In order to achieve finer-grained analysis, \citet{song2018personal} proposed a taxonomy-guided multi-task learning model to detect what personal aspects of online users are disclosed in online posts. They also constructed a dataset of privacy-disclosing tweets covering 32 privacy-relevant personal aspects. Similarly, \citet{Chen2020GrHA} proposed GrHA, a fine-grained privacy detection network, to improve the performance of the model proposed in~\citep{song2018personal}. The above two proposed methods aim to improve the prediction results by considering label co-occurrences, but they did not consider label-to-text correlations explicitly.

\subsection{Multi-label Text Classification}

Traditional machine learning methods~\citep{kumar2012learning,jacob2008clustered} have been widely used to deal with MLTC tasks. \citet{kumar2012learning} proposed the GO-MTL model by using grouping and overlap mechanism to enhance the semantic correlations in MLTC tasks. Likewise, \citet{jacob2008clustered} studied the clustered multi-task learning to deal with MLTC tasks. Although these machine learning methods utilize multiple hand-crafted features to enhance the semantic representations in MLTC tasks, they overlook deep semantic features among input text and multi labels.

Nowadays, researchers have made great progress on the deep learning technology. Therefore deep models such as CNN~\citep{Liu_CNN, 2016Improved, kim2014convolutional} and RNN~\citep{2016Recurrent,Chen_RNN} have been used to implement end-to-end MLTC tasks. In more recent studies, researchers have also proposed to use attention mechanisms such as DocBERT~\citep{2019DocBERT} and other methods such as SGM~\citep{SGM_2018} and LSAN~\citep{LSAN_2019} to consider the label-to-text correlation in the MLTC problem. \cite{2019DocBERT} proposed DocBERT model as a much simpler BERT model with competitive accuracy at a far more modest computational cost in terms of MLTC tasks. \citet{SGM_2018} considered how to address the MLTC problem by capturing the correlations between labels as well as the most informative words automatically when predicting different labels. \citet{LSAN_2019} used self-attention and label-attention for better representations of input text in MLTC tasks. Label co-occurrences are a vital source of information when dealing with the MLTC problem. More specifically, some labels often appear with other labels due to the semantic relation. However, most existing methods focus only on optimizing the process of feature extraction, but do not consider label co-occurrences. By utilizing the GCN model, \citet{LDGN} proposed LDGN (label-specific dual graph neural network) to improve the MLTC representations by including label co-occurrences. Although they considered label co-occurrences to a certain extent, their method has some limitations in the process of combination with the feature exaction module, for their model's usage of the GCN only attempts to optimize the text representation of the model with label co-occurrences yet ignores diversity of the text representation and the labels' guidance on fusing different feature vectors.

\section{Proposed Method}
\label{sec:proposed_method}

\begin{figure}[!htb]
\centering
\includegraphics[width=\linewidth]{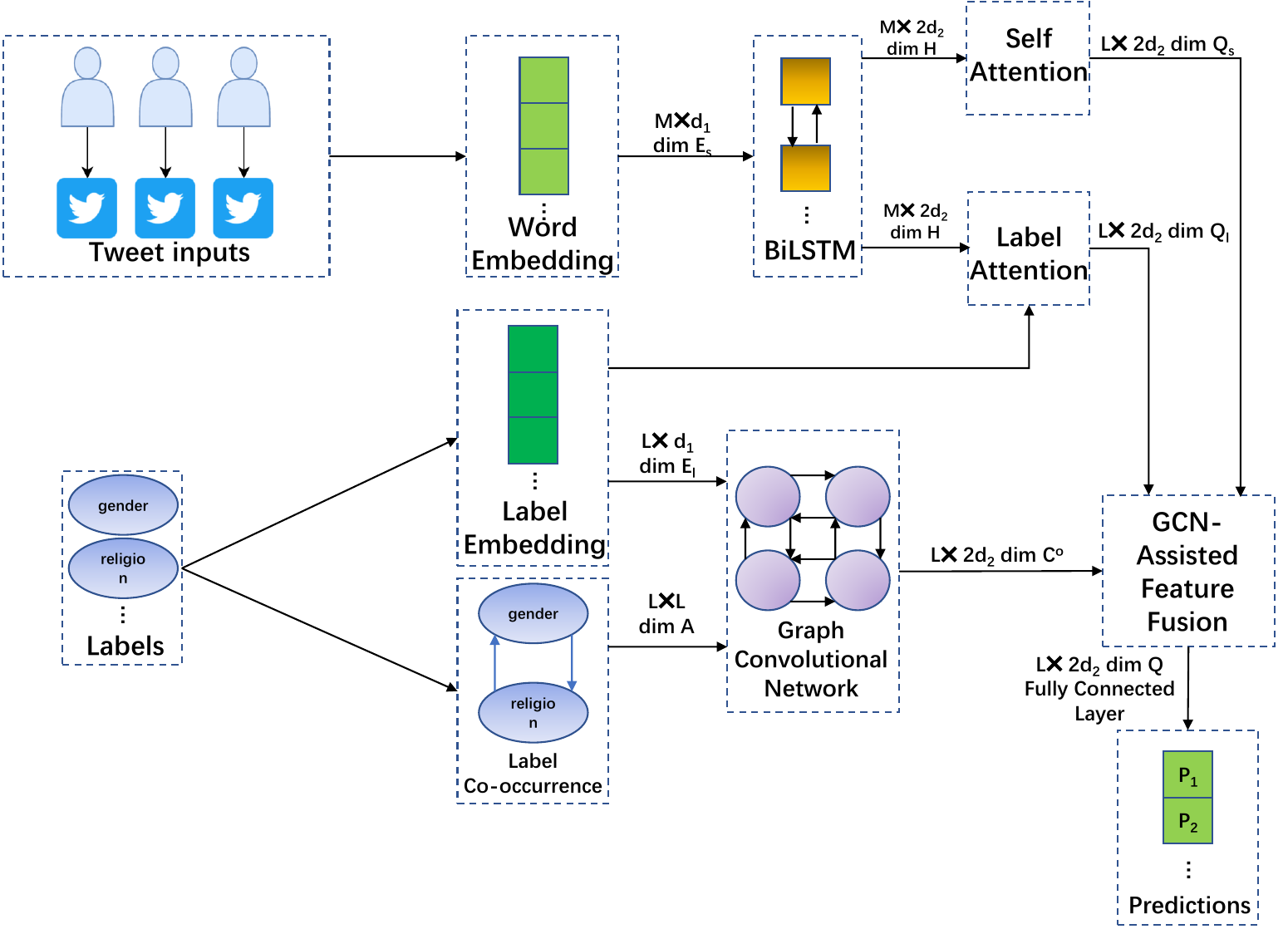}
\caption{The architecture of the proposed network.}
\label{fig:proposed_network}
\end{figure}

In this section, we introduce the GCN-based double attention network, as shown in Fig.~\ref{fig:proposed_network}. The network includes four major components: 1) an input text feature encoder that transforms the input text into word-level semantic vectors; 2) a double-attention text representation component that enhances the important word representations of the text combining both text information and label information; 3) a GCN-assisted feature fusion mechanism that utilizes the label-to-label correlation acquired by GCN to guide the double-attention information fusion process; and 4) a label probability output component that predicts the probabilities of various privacy-relevant labels.

\subsection{Problem Formulation}

Let $\mathbb{D}=\left\{(x_i, y_i)\right\}_{i=1}^N$ denote the set of texts, where $x_i$ represents the input texts and $y_i\in\{0,1\}^L$ represents its corresponding labels. Here, $L$ denotes the total number of privacy-relevant labels. The target of the proposed method in this paper is to learn the output probability of each label from the input text, in order to match the most relevant labels.

\subsection{Input Text Feature Encoder}

Given a text $x_i$ containing $M$ words $(x_i=\left\{w_{i1}, w_{i2}, \cdots, w_{iM}\right\})$, the word2vec method~\citep{le2014distributed} is adopted to obtain the embedding vector based on the input, which is denoted as $\mathbf{E_s} \in \mathbb{R}^{M \times d_1}$, where $d_1$ denotes the embedding dimension.

For fair comparisons, we used the same feature extraction structure, bidirectional long short-term memory (BiLSTM)~\citep{zhou2016text}, as the baseline models~\citep{Chen2020GrHA, LSAN_2019, LDGN} used, to get the embedding. We adopt the BiLSTM model to process the embedded vector. The formula is as follows:
\begin{equation}
\begin{aligned}
&\mathbf{H}=\left\{\overrightarrow{\mathbf{H}}_r, \overleftarrow{\mathbf{H}}_l\right\},\\
\end{aligned}
\end{equation}
where $\overrightarrow{\mathbf{H}}_r, \overleftarrow{\mathbf{H}}_r \in \mathbb{R}^{M\times d_2}$ represent the forward and backward text representations, respectively. The whole text can be represented as $\mathbf{H} \in \mathbb{R}^{M\times 2d_2}$.

\subsection{Double-Attention Text Representation}

We use a double attention mechanism to generate text- and label-specific representations from the output of the BiLSTM. A self-attention model is adopted to capture the long-term dependence of words in $\mathbf{H}$. Meanwhile, to extract the text attention from the corresponding labels, a label-specific attention model is used as the supplementary information.

\subsubsection{Self-Attention Model}

Self-attention models have shown their considerable merits on assessing the importance of word representations. Therefore, we adopt a self-attention mechanism~\citep{lin2017structured} to reinforce the semantic representation of the text based on the word-to-word correlations. Different from traditional self-attention algorithms, the self-attention sentence embedding algorithm~\citep{lin2017structured} uses multiple hops of attention calculated from the LSTM outputs $\mathbf{H}$ to focus on different aspects of the meanings of the sentence. Since the output labels have the dimensionality of $L$, we take the self-attention weights with $L$ dimensions to reflect the effects of $L$ labels to $M$ words. The calculation of attention weights can be described as follows:
\begin{equation}
\mathbf{A}_s=\operatorname{softmax}\left(\mathbf{W}_{s2} \tanh \left(\mathbf{W}_{s1} \mathbf{H}^T\right)\right),
\end{equation}
where $\mathbf{A}_s \in \mathbb{R}^{L \times M}$ are the self attention weights that indicate the effect of each word to each label. $\mathbf{W}_{s1} \in \mathbb{R}^{d_3 \times 2d_2}, \mathbf{W}_{s2} \in \mathbb{R}^{L \times d_3}$ are the parameters to be trained. Then, the attention weights are utilized to update the text representation:
\begin{equation}
\mathbf{Q}_s=\mathbf{A}_s \times \mathbf{H}^T.
\end{equation}

\subsubsection{Label-Attention Model}

Apart from obtaining text attention from the text itself, the label-attention model~\citep{LSAN_2019} is adopted to extract text attention from the corresponding labels. The labels' semantic information is acquired with the word2vec method, which is denoted as $\mathbf{E}_l \in \mathbb{R}^{L \times d_1}$.

To capture a better semantic representation with the guidance of output labels, the label-attention mechanism computes the attention weights by calculating the relationship between the labels and the text as follows $\mathbf{A}_l=\mathbf{E}_l \times \mathbf{H}^T$, where $\mathbf{A}_l \in \mathbb{R}^{L \times M}$ are the label-specific attention weights that indicate the effect of each word to each label. With the weight matrix, the label-specific attention weights are utilized to enhance the label-aware information in the text semantic representation $\mathbf{Q}_l=\mathbf{A}_l \times \mathbf{H}^T$.

\subsection{GCN-Assisted Feature Fusion}

In this section, the GCN-assisted feature fusion mechanism is described to construct comprehensive text representations with the guidance of the label-to-label correlation.

We use a GCN framework to extract a label-to-label correlation matrix. With the guidance of the correlation matrix, we enhance the text representations by utilizing the proposed compensation coefficients to implement the algorithm of feature fusion.

\subsubsection{GCN-based Label-to-Label Correlation Extraction}

The graph convolutional networks (GCNs)~\citep{kipf2016semi} were proposed to get a better understanding of the relationship of nodes in a graph. A GCN uses an adjacency matrix to characterize the graph structure and a convolutional network to capture the correlations among different nodes, with an output of a correlation matrix. In our work, we aim to extract the label co-occurrence through a GCN. The label co-occurrence refers to the simultaneous occurrence of two or more labels in the same text. For example, considering the two labels ``Salary" and ``Occupation", their probability of co-occurrence is high due to their semantic relation (i.e., an occupation is normally associated with a salary). Therefore, we utilize the GCN to transform such label-to-label relationships (inferred from label co-occurrences and their semantic relationships) into mathematical representations.

\begin{figure}[!htb]
\centering
\begin{tikzpicture}[node distance=1cm, auto]  
\tikzset{
    mynode/.style={rectangle,rounded corners,draw=black, top color=white, bottom color=yellow!50,very thick, inner sep=0.5em, minimum size=0.5em, text centered,text width=6em},
    myarrow/.style={->, >=latex', shorten >=1pt, thick},
    mylabel/.style={text width=7em, text centered} 
}  

\node[mynode] (labela) {Label a Embedding};  
\node[below=4cm of labela] (dummy) {}; 
\node[mynode, left=2cm of dummy] (labelb) {Label b Embedding};  
\node[mynode, right=2cm of dummy] (labelc) {Label c Embedding};
\node[mylabel, below left=2cm of labela] (a_to_b) {$p_a^b=\#(a,b)/\#(a)$};  
\node[mylabel, right=0.1cm of a_to_b] (b_to_a) {$p_b^a=\#(b,a)/\#(b)$};  
\node[mylabel, below right=2cm of labela] (a_to_c) {$p_a^c=\#(a,c)/\#(a)$};  
\node[mylabel, left=0.1cm of a_to_c] (c_to_a) {$p_c^a=\#(c,a)/\#(c)$};  
\node[mylabel, below=2.8cm of labela] (b_to_c) {$p_b^c=\#(b,c)/\#(b)$};  
\node[mylabel, below=4.2cm of labela] (c_to_b) {$p_c^b=\#(c,b)/\#(c)$};  

\draw[myarrow]
(3.4,-4) to  (0.6,-0.6);
\draw[myarrow]
(1.1,-0.6) to (3.9,-4);
\draw[myarrow]
(-3.4,-4) to  (-0.6,-0.6);
\draw[myarrow]
(-1.1,-0.6) to (-3.9,-4);
\draw[myarrow]
(-2.1,-4.5) to (2.1,-4.5);
\draw[myarrow]
(2.1,-4.8) to (-2.1,-4.8);
\end{tikzpicture} 
\caption{Construction of the initial weighted label graph.} 
\label{fig:gcn_explanation}
\end{figure}
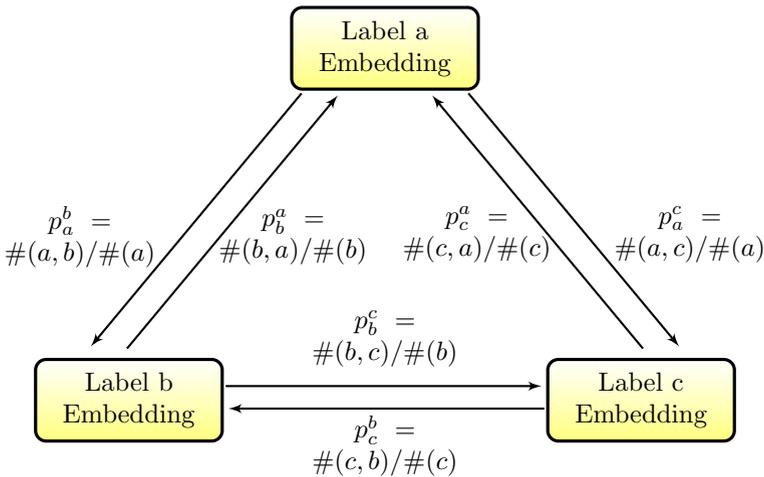

As Fig.~\ref{fig:gcn_explanation} shows, the output labels are represented as a weighted label graph $(\mathbf{V},\mathbf{E})$, where each node represents a label embedding and each edge's weight refers to the two adjacent labels' co-occurrence frequency. More specifically, each node is initialized to be the embedded vector of the corresponding label and each edge weight is calculated to be the co-occurrence frequency of the two labels representing the two adjacent nodes based on information in the training set. In Fig.~\ref{fig:gcn_explanation}, the symbol $\#$ represents the the number of occurrences. For example, $\#(a)$ represents the number of tweets with the label $a$ in the training set and $\#(a,b)$ represents the number of tweets with both labels $a$ and $b$ in the training set. We use $\mathbf{P}$ to represent the initial co-occurrence adjacent matrix. According to~\cite{Chen2020GrHA}, considering the noisy co-occurrence caused by the sparse real-world dataset, the initial co-occurrence adjacent matrix $\mathbf{P}$ should be binarized and revised as follow:
\begin{equation}
a_j^k= \begin{cases}\frac{u}{\sum_{x=1}^L p_j^k}, & \text{ if } j \neq k, \\ 1-u, & \text{ if } j=k,\end{cases}
\end{equation}
where $p_j^k$ represents the co-occurrence frequency of label $j$ to label $k$ and $a_j^k$ represents the revised co-occurrence frequency. $u$ represents the trade-off parameter that balances the weights between the label itself and its correlated labels. We use $\mathbf{A}$ to represent the revised adjacency matrix. In our work, we use the same revised adjacency matrix as~\cite{Chen2020GrHA} did. The trade-off parameter is set to 0.2.

Then, a GCN is adopted to update the label-to-label correlation representations from the previous representations and the adjacency matrix containing co-occurrence probabilities. The GCN propagation is calculated as follows:
\begin{equation}
\mathbf{C}^{(l+1)}=\sigma\left(\mathbf{AC}^{(l)} \mathbf{W}^{(l)}_g\right),
\end{equation}
where $\mathbf{C}^{(l)} \in \mathbb{R}^{L \times d_4^{(l)}}$ represents the input label-to-label correlation representations for the $l$-th GCN layer, $\sigma$ denotes the activation function (LeakyReLU is adopted here), $\mathbf{A}$ is the revised adjacency matrix, and $\mathbf{W}^{(l)}_g \in \mathbb{R}^{d_4^{(l)} \times d_4^{(l+1)}}$ denotes the transformation matrix to be learned for the $l$-th layer.

Our GCN  contains two layers. As a result, the second layer's embedding size adopts $2d_2$ to align the dimension of the output from the double-attention model. Thus the correlation matrix is obtained from the output of the second layer, which is denoted as $\mathbf{C}^{\text{out}}\in \mathbb{R}^{L \times 2d_2}$.

\subsubsection{Feature Fusion Guided by Label-to-Label Correlation}

As mentioned above, we obtain the text representations including the text semantic information (from self-attention) and the label-to-text correlation (from label-attention), and represent the label-to-label correlation through a GCN. The text semantic information uses the self-attention mechanism to enhance the weight of key words or phrases based on the inputting text semantics itself. Meanwhile, the label-to-text correlation provides the improved text representations through the label-attention mechanism, which is based on the labels' semantic representations. Therefore, these two text representations shuffle the word weights of the input texts to enhance their key parts. However, they are based on different semantic information (the text itself and the labels' semantics) and the enhanced parts are different. Therefore, it is important and necessary to fuse these two representations in order to get a more comprehensive semantic representations. To this end, we propose a cross-attention model that utilizes the label-to-label correlation matrix to guide the fusion of output features from the double-attention model. The experimental results demonstrated the superiority of our model compared to other state-of-the-art methods.

Our method aims to enhance the weak part of the representations in the output from different attention models and utilize the label-to-label correlation to fuse such output features better. More specifically, the output from the self-attention mechanism enhances the key words or phrases according to the context semantics of inputting texts yet lacks the representation enhancement from label-text correlation features, while the output from the label-attention mechanism enhances the key words or phrases according to the label semantics yet lacks the representation enhancement from text semantic features. Therefore, with the guidance of a GCN, we aim to acquire the complementary feature vectors of these two representations. We use the proposed compensation coefficients guided by the GCN to quantify the extent of the compensation above. First, we calculate the cross-attention weights, denoted by $\mathbf{W}_l, \mathbf{W}_s \in \mathbb{R}^L$, which indicate the compensation coefficients of each representation. The model's output can be described as follows:
\begin{equation}
\begin{aligned}
& \mathbf{W}_l=f\left(\mathbf{C}^{\text{out}} \mathbf{Q}_s^T \mathbf{W}_{a1}\right),\\
& \mathbf{W}_s=f\left(\mathbf{C}^{\text{out}} \mathbf{Q}_l^T \mathbf{W}_{a2}\right),\\
& \mathbf{W}_l+\mathbf{W}_s=\mathbf{1},
\end{aligned}
\end{equation}
where $\mathbf{W}_{a1},\mathbf{W}_{a2} \in \mathbb{R}^L$ are parameters to be trained, $f$ represents the sigmoid function, the third equation is to let $\mathbf{W}_l$ and $\mathbf{W}_s$ satisfy the normalization constraint, and $\mathbf{1}$ represent an all-one vector. Then, according to the compensation coefficients, the $i$-th label based final text representation can be obtained as $\mathbf{Q}_i=\mathbf{W}_{li}\mathbf{Q}_{li}+\mathbf{W}_{si}\mathbf{Q}_{si}$. The final text representation output by the proposed model is $\mathbf{Q}=\{\mathbf{Q}_i\}_{i=1}^L \in \mathbb{R}^{L \times 2d_2}$.

\subsection{Label Probability Prediction}

After obtaining the fused text representation, we  feed $\mathbf{Q}$ into a fully connected layer for the label probability prediction to produce the prediction result $\hat{y}=f\left(\mathbf{Q}\mathbf{W}_o\right)$, where $f$ represents the sigmoid function and $\mathbf{W}_o \in \mathbb{R}^{2d_2}$ are the parameters to be trained.

After comparing the predicted labels $\hat{y}$ with the ground-truth $y \in\{0,1\}^L$, the proposed model is trained with the cross entropy loss as follows:
\begin{equation}
\mathcal{L}=\sum_{l=1}^L y_l \log \left(\hat{y}_l\right)+\left(1-y_l\right) \log \left(1-\hat{y}_l\right).
\end{equation}

\section{Experimental Results}
\label{sec:experimental_results}

To evaluate our proposed model, we conducted numerous experiments on a public dataset of privacy-disclosing tweets and compared the performance of our model with selected state-of-the-art methods in terms of key performance metrics. Furthermore, we verified the effect of each component in our model with corresponding ablation tests and component analysis. Finally, we used our proposed model to test some concrete tweet examples to demonstrate the practicability of the proposed model.

\subsection{Experimental Setup}

\subsubsection{Dataset Used}

\begin{figure}[!h]
\centering
\includegraphics[width=\linewidth]{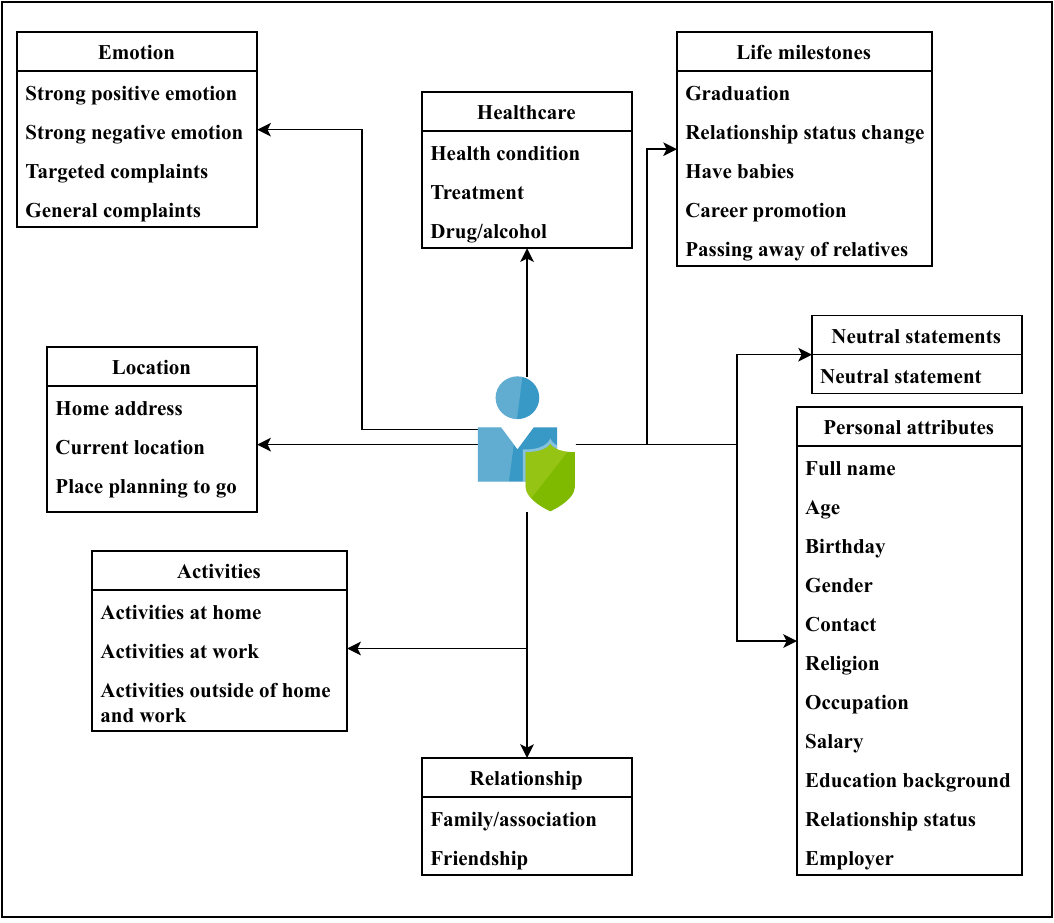}
\caption{Illustration of 32 categories of privacy used in our experiments.}
\label{fig:privacy_categories}
\end{figure}

We evaluated our proposed model on the public dataset of privacy-disclosing tweets introduced in~\citep{song2018personal}, which includes 11,368 tweets each annotated with one or more privacy-relevant labels representing 32 privacy-oriented personal aspects. Fig.~\ref{fig:privacy_categories} illustrates 32 categories of privacy in the dataset specifically. In the dataset, the personal privacy is firstly divided into eight groups, including ``Healthcare'', ``Life milestones'', ``Personal attributes'', ``Relationship'', ``Activities'', ``Location'', ``Emotion'' and ``Neutral statements''. The first seven groups represent seven general privacy groups and the last group ``Neutral statements'' represents those tweets that do not disclose any category of privacy. These eight groups make a higher-level categorization of privacy-related information, which covers most of personal privacy disclosures we can observe in the real world. Furthermore, the eight privacy groups are subdivided into 32 finer-grained privacy categories, which show different types of privacy-related information more specifically. Our experiments are based on 32 privacy-oriented personal aspects and each label represents one privacy-oriented personal aspect. To the best of our knowledge, no any other public datasets offer a comparable level of richness and comprehensiveness considering the size of the dataset and the richness of privacy-oriented personal aspects. Table~\ref{tab:average_labels} shows the number of tweets with a specific quantity of unique personal aspects. An average tweet is annotated with 1.31 personal aspects.

\begin{table}[!h]
\centering
\caption{The number of tweets with a specific quantity of unique personal aspects, as annotated in the Twitter dataset.}
\label{tab:average_labels}
\begin{tabular}{c *{7}{c}}
\toprule
\textbf{\#(Personal Aspects)} & 1 & 2 & 3 & 4 & 5 & $>5$\\
\midrule
\textbf{\#(Tweets)} & 8,546 & 2,215 & 533 & 65 & 9 & 0\\
Percentage & 75.2\% & 19.5\% & 4.7\% & 0.57\% & 0.079\% & 0\%\\
\bottomrule
\end{tabular}
\end{table}

\subsubsection{Evaluation Metrics}

Following the settings of previous work~\citep{Chen2020GrHA}, we use average precision (Avg-prec), one-error (One-err), precision at top K (P@K) and S@K for performance evaluation, which are explained as follows:

\textbf{Average Precision} (Avg-pre) \quad Average precision evaluates the overall precision of the input texts over the ranking list of labels according to the ground truth~\citep{2013Multi}.

\textbf{One-Error} (One-err) \quad One-error represents the mean possibility that the first prediction of the personal aspects does not conform to the ground truth~\citep{zhang2007multi}.

\textbf{P@K} \quad P@K refers to the average precision of label predictions among the top K recommended results.

\textbf{S@K} \quad S@K refers to the mean probability that a correct personal aspect is captured within the top K recommended results~\citep{song2018personal}.

\subsubsection{Parameter Settings}

For fair comparisons, we split the dataset in our experiments in the same way as in previous work~\citep{song2018personal, Chen2020GrHA}. The experimental results were obtained through the 10-fold cross-validation.

We split the training set into a training subset and a validation subset whose ratio is 8:1. We selected the best parameter configuration based on the validation performance, i.e., the hyper-parameter fine-tuning was completed based on evaluation metrics calculated from the validation subset. To obtain the word embedding and label embedding, we utilized the word2vec method to convert texts into 300 dimensional vectors, which means $d_1=300$. The BiLSTM hidden dimension is set as $d_2=300$. The hyper-parameter corresponding to the self-attention mechanism is set as $d_3=200$. Furthermore, our model's GCN uses a 2-layer model with the hidden dimension of 450. The batch size searched are 16, 32, 64, and 128, and the learning rate searched are 0.1, 0.01, 0.001, and 0.0001. According to the validation performance, we took 64 as the batch size, and used the Adam optimizer~\citep{2014Adam} to minimize the loss with the initial learning rate of 0.001. We use the Floating-Point Operations (FLOPs) and Multiply-Accumulates (MACs) to measure the computational complexity of the proposed model. The experimental results indicate that the FLOPs of the proposed model is 12.61G and the MACs of the proposed model is 1.59M.

\subsection{Baseline Models}

First, we compared our proposed model with several methods for predicting privacy disclosures in online posts, including five shallow learning methods and four deep learning methods. To further demonstrate our proposed method's performance, we compared it with two recent state-of-the-art MLTC models. Therefore, we used the following eleven models as baselines.

\begin{itemize}
\item \textbf{SVM}~\citep{1995Support}: A classical machine learning model that concatenates the privacy-oriented features into a single vector and learns each personal aspect individually.

\item \textbf{MTL-Lasso}~\citep{tibshirani1996regression}: A multi-task learning method (MTL) with Lasso which implements the $l_1$-penalization to the regression objective function.

\item \textbf{GO-MTL}~\citep{kumar2012learning}: A model using grouping and overlap mechanism to learn the semantic correlations among personal aspects.

\item \textbf{CMTL}~\citep{jacob2008clustered}: The clustered multi-task learning (CMTL) which assumes personal aspects can be clustered into several groups and each group can be learned together.

\item \textbf{TOKEN}~\citep{song2018personal}: The latent group MTL that utilizes the pre-defined personal aspect taxonomy to learn the group-sharing and aspect-specific latent features of personal aspects simultaneously.

\item \textbf{TextRNN}~\citep{giles1994dynamic}: A RNN-based model which uses RNN and logistic regression for privacy disclosure detection.

\item \textbf{TextCNN}~\citep{kim2014convolutional}: A CNN-based model which also uses CNN and logistic regression (similar to TextRNN) for privacy disclosure detection.

\item \textbf{D-TOKEN}~\citep{song2018personal}: An end-to-end model as an extension of TOKEN, which replaces the hand-crafted features by representation automatically learned by hierarchical attentive network (HAN).

\item \textbf{GrHA}~\citep{Chen2020GrHA}: A HAN-based privacy detection model which uses graph-regularization mechanism to enhance label co-occurrences representations.

\item \textbf{LSAN}~\citep{LSAN_2019}: A label-specific attention network model based on self-attention and label-attention mechanism.

\item \textbf{LDGN}~\citep{LDGN}: A label-specific dual graph network model which contains label-attention and dual graph neural network.
\end{itemize}

\subsection{Experimental Results and Discussion}

\begin{table}[!htb]
\centering
\caption{Performance comparisons with selected state-of-the-art methods on the dataset used. Partial experimental results of baseline models are directly extracted from~\cite{Chen2020GrHA}.}
\label{tab:experiment_results}
\resizebox{\textwidth}{!}{
\begin{tabular}{cc cccccccc}
\toprule
\multicolumn{2}{c}{\textbf{Model}} & Avg-pre & One-err & S@1 & S@3 & S@5 & P@1 & P@3 & P@5\\
\midrule
\multirow{5}{*}{Shallow} &
SVM & 52.91\% & 69.35\% & 30.65\% & 72.98\% & 80.47\% & 30.65\% & 26.33\% & 18.47\% \\
& MTL-Lasso & 58.00\% & 56.09\% & 43.91\% & 73.18\% & 82.11\% & 43.91\% & 27.38\% & 19.31\% \\
& GO-MTL & 58.68\% & 56.02\% & 43.98\% & 74.24\% & 83.92\% & 43.98\% & 27.65\% & 19.78\% \\
& CMTL & 58.99\% & 55.84\% & 44.16\% & 74.41\% & 83.30\% & 44.16\% & 27.81\% & 19.63\% \\
& TOKEN & 59.05\% & 55.96\% & 44.04\% & 74.72\% & 84.34\% & 44.04\% & 27.96\% & 19.92\%\\
\midrule
\multirow{7}{*}{Deep} &
TextRNN & 61.84\% & 49.61\% & 50.39\% & 74.50\% & 82.64\% & 50.39\% & 28.67\% & 19.90\% \\
& TextCNN & 69.31\% & 39.99\% & 60.01\% & 81.97\% & 88.40\% & 60.01\% & 31.77\% & 21.44\% \\
& D-TOKEN & 69.43\% & 39.96\% & 60.04\% & 83.39\% & 89.35\% & 60.04\% & 32.15\% & 21.56\% \\
& GrHA & 71.44\% & 39.17\% & 60.83\% & 85.83\% & 92.20\% & 60.83\% & 33.51\% & 22.60\% \\
& LSAN & 73.62\% & 37.40\% & 62.60\% & 88.32\% & 94.25\% & 62.60\% & 34.98\% & 23.50\% \\
& LDGN & 73.94\% & 36.85\% & 63.15\% & 88.36\% & 94.45\% & 63.15\% & 35.03\% & 23.57\%\\
\midrule
& \textbf{Our Model} & \textbf{74.30\%} & \textbf{36.35\%} & \textbf{63.65\%} & \textbf{89.08\%} & \textbf{95.00\%} & \textbf{63.65\%} & \textbf{35.23\%} & \textbf{23.60\%}\\
\bottomrule
\end{tabular}
}
\end{table}

Table~\ref{tab:experiment_results} shows the performance metrics of all the compared methods, all based on the same dataset. For LSAN and LDGN, the two most recent baseline models, the experimental results were obtained from our own experiments. For other baseline models, the performance figures were taken from~\citep{Chen2020GrHA}, which were obtained using the same dataset and experimental settings as we used. The results show that our method outperformed all other baseline models, proving the effectiveness of the double-attention mechanism and the GCN-assisted feature fusion mechanism.

For all the evaluated models, deep learning methods are proved to access better results than shallow learning methods, which shows the importance of neural network on extracting text's features. Among all the deep models, TextRNN, TextCNN, D-TOKEN are less effective because those models only focus on the features of the text and ignore the relationship between text and labels. GrHA and LSAN improve the results to a certain extent, on account for using the attention mechanism to extract the texts' correlation. However GrHA ignores the label-to-text correlation and directly utilizes the GCN to introduce label co-occurrences rather than assisting the feature fusion process. LSAN does not consider the impact of labels' co-occurrence, which causes the adverse effects on final results. LDGN uses label-attention and dual graph neural network to make up the deficiency of co-occurrence for labels. However by comparing with LDGN and our proposed model, the latter outperforms because its methods for processing label-to-label correlation is based on the GCN-assisted feature fusion mechanism, which uses the compensation coefficients to guide the fusion of text representations, while LDGN only uses the dot product operation.

In conclusion, the proposed network outperforms shallow models, deep embedding models, label attention based models. The improvement of the proposed model demonstrates the effectiveness of the double attention mechanism and the proposed GCN-assisted feature fusion mechanism.

\subsection{Ablation Tests}

\begin{table}[!htb]
\centering
\caption{Ablation tests of our proposed method using six different possible combinations of the three key components.}
\label{tab:ablation_test}
\resizebox{\textwidth}{!}{
\begin{tabular}{c ccc cccccccc}
\toprule
\textbf{Model} & Avg-pre & One-err & S@1 & S@3 & S@5 & P@1 & P@3 & P@5\\
\midrule
\textbf{S} & 72.37\% & 37.74\% & 62.26\% & 85.72\% & 92.52\% & 62.26\% & 33.90\% & 22.88\%\\
\textbf{L} & 73.25\% & 37.66\% & 62.34\% & 87.90\% & 94.22\% & 62.34\% & 34.75\% & 23.31\%\\
\textbf{SG} & 72.39\% & 37.73\% & 62.27\% & 85.71\% & 92.53\% & 62.27\% & 33.91\% & 22.90\%\\
\textbf{LG} & 73.32\% & 37.62\% & 62.38\% & 87.95\% & 94.20\% & 62.38\% & 34.77\% & 23.32\%\\
\textbf{SL} & 73.62\% & 37.40\% & 62.60\% & 88.32\% & 94.25\% & 62.60\% & 34.98\% & 23.50\%\\
\midrule
\textbf{SLG} & \textbf{74.30\%} & \textbf{36.35\%} & \textbf{63.65\%} & \textbf{89.08\%} & \textbf{95.00\%} & \textbf{63.65\%} & \textbf{35.23\%} & \textbf{23.60\%}\\
\bottomrule
\end{tabular}
}
\end{table}

A series of ablation tests were conducted to show the contribution of each module in the proposed network. Since the proposed model has three functional modules, the self-attention module (S), the label-attention module (L) and the GCN-assisted feature fusion module (G), in the ablation tests, we experimented all six possible combinations of the three modules: S, L, SL (which is effectively LSAN), SG, LG, and SLG (which is our model). Note that G cannot be used alone.

As Table~\ref{tab:ablation_test} presents, Model LG outperformed Model L while Model SG outperformed Model S, which shows the function of the GCN-assisted feature fusion module. Meanwhile aforementioned improvement is slight, which indicates that the GCN-assisted feature fusion module can exhibit its maximum function only with double attention mechanism. Model SL performed better than Models LG and SG, which indicates that the text representation is still the core process of the privacy MLTC. Model LG outperformed Model SG, which demonstrates that the label-attention mechanism can capture the feature of texts and labels more effectively and more accurately than the self-attention mechanism. Our proposed model (SLG) gained the best performance for all metrics, showing that combining all the three sources of information is indeed effective.

\subsection{Component Analysis}

To further illustrate the performance of the proposed model, we conducted some further analysis for each component of our proposed model and present several samples selected from the privacy dataset we used.

\subsubsection{Label Attention Weights}

\begin{figure}[!htb]
\centering
\includegraphics[width=\linewidth]{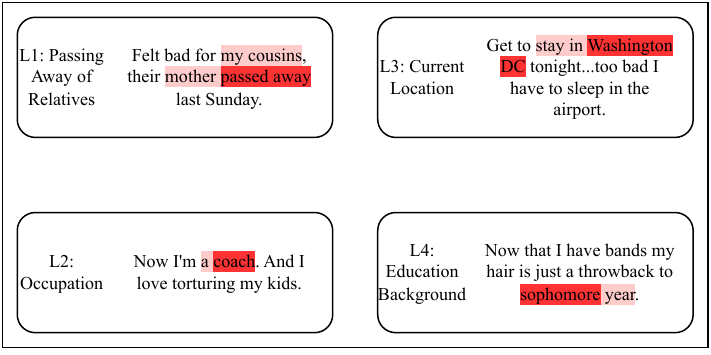}
\caption{The visualization of label attention weights.}
\label{fig:attention_case}
\end{figure}

We can use heat maps to show the label attention weights. For several test samples from the test set of our dataset, such a heat map is shown in Fig.~\ref{fig:attention_case}. The brightness of the red bar represents the label attention weight of each word (darker = larger weight), according to the double-attention mechanism. For example, the more significant words for the label ``Occupation'' are ``a coach''. For the label ``Current Location'', the label attention mechanism focuses on names of places such as ``Washington DC''. Generally speaking, the label attention mechanism is capable of extracting important information in the input text and benefiting the subsequent classification module.

\subsubsection{GCN-Assisted Feature Fusion}

\begin{table}[!htb]
\centering
\caption{The performance comparison of models based on our GCN-based module and three dot-product-based modules.}
\label{tab:gcn_module_test}
\resizebox{\textwidth}{!}{
\begin{tabular}{c cccccccc}
\toprule
\textbf{Model} & Avg-pre & One-err & S@1 & S@3 & S@5 & P@1 & P@3 & P@5\\
\midrule
\textbf{Dot-S} & 72.45\% & 38.68\% & 61.32\% & 87.19\% & 93.59\% & 61.32\% & 34.11\% & 23.20\%\\
\textbf{Dot-L} & 73.01\% & 37.72\% & 62.28\% & 87.30\% & 93.45\% & 62.28\% & 34.59\% & 23.22\%\\
\textbf{Dot-SL} & 73.57\% & 37.18\% & 62.82\% & 88.08\% & 93.85\% & 62.82\% & 34.91\% & 23.25\%\\
\midrule
\textbf{Our Model} & \textbf{74.30\%} & \textbf{36.35\%} & \textbf{63.65\%} & \textbf{89.08\%} & \textbf{95.00\%} & \textbf{63.65\%} & \textbf{35.23\%} & \textbf{23.60\%} \\
\bottomrule
\end{tabular}
}
\end{table}

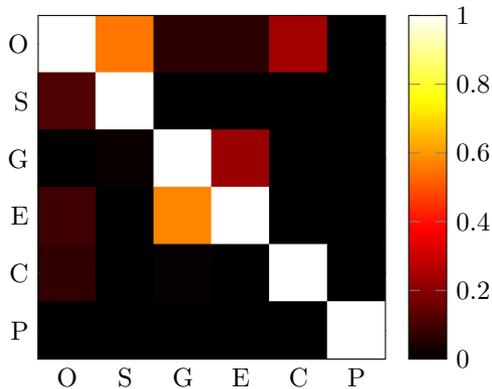
\begin{figure}[ht]
\centering
\pgfplotsset{
    heatmap axis style/.style={
        width=0.6\linewidth,
        enlargelimits=true,
        colorbar=true,
        colormap/hot2,
        symbolic x coords={O,S,G,E,C,P},
        symbolic y coords={O,S,G,E,C,P},
        ymax=P,
        ymin=O,
        xmax=P,
        xmin=O,
        xtick=data,
        ytick=data,
        major tick length=0pt,
        minor tick length=0pt,
        axis equal image
    }
}

\begin{tikzpicture}
\begin{axis}[heatmap axis style]
\addplot [
matrix plot,
mesh/cols = 6,
point meta=explicit,
] coordinates {
(O,O) [1.0](O,S) [0.1162](O,G) [0.001](O,E) [0.0917](O,C) [0.0703](O,P) [0]

(S,O) [0.5507](S,S) [1.0](S,G) [0.0145](S,E) [0.001](S,C) [0.001](S,P) [0]

(G,O) [0.0638](G,S) [0.001](G,G) [1.0](G,E) [0.5691](G,C) [0.01](G,P) [0]

(E,O) [0.0634](E,S) [0.001](E,G) [0.2267](E,E) [1.0](E,C) [0.001](E,P) [0]

(C,O) [0.2396](C,S) [0.001](C,G) [0.001](C,E) [0.001](C,C) [1.0](C,P) [0]

(P,O) [0](P,S) [0](P,G) [0](P,E) [0](P,C) [0](P,P) [1.0]
};
\end{axis}
\end{tikzpicture}
\caption{The visualization of labels' co-occurrence adjacency matrix (O = ``Occupation'', S = ``Salary'', G = ``Graduation'', E = ``Education'', C = ``Career promotion'', and P = ``Passing Away of Relatives'').}
\label{fig:gcn_hm}
\end{figure}

To show the effectiveness of the GCN-assisted feature fusion visually, we can also use a heat map representing label co-occurrences. One example is given in Fig.~\ref{fig:gcn_hm}, which shows that the label ``Occupation'' correlates highly with the label ``Salary'', and the label ``Graduation'' correlates highly with the label ``Education''. Besides, the label ``Education'' correlates with the label ``Graduation'' to some extent. On the other hand, the label ``Passing Away of Relatives'' is almost irrelevant to other labels due to their lack of semantic connections. The example demonstrates that the GCN-based model can extract label-to-label relationships with the graph structure quite effectively.

To provide further evidence of the effectiveness of our GCN-based method, we also compared the performance of two groups of distinct GCN-based modules: our proposed GCN-assisted feature fusion module and the more common dot-product-based GCN modules. For the latter, we considered three possible modules: \textbf{Dot-S} -- the dot-product-based model with self attention only, \textbf{Dot-L} -- the dot-product-based model with label attention only, and \textbf{Dot-SL} -- the dot-product-based model with double attention. The comparison results are shown in Table~\ref{tab:gcn_module_test}, which shows that our proposed GCN-based module outperformed all the other three dot-product-based modules. Compared with the dot-product-based modules, our module utilizes the label-to-label correlation matrix to guide the fusion of the output from the double-attention network, which can gain a better text representation.

\subsubsection{Number of GCN Layers}

\begin{table}[htb]
\centering
\caption{The evaluation of performance on different numbers of GCN layers.}
\label{tab:gcn_layer_test}
\resizebox{\textwidth}{!}{
\begin{tabular}{c cccccccc}
\toprule
\textbf{Model} & Avg-pre & One-err & S@1 & S@3 & S@5 & P@1 & P@3 & P@5 \\
\midrule
\textbf{GCN-1} & 73.78\% & 37.46\% & 62.54\% & 88.10\% & 94.85\% & 62.54\% & 34.97\% & 23.58\% \\
\textbf{GCN-2} & \textbf{74.30\%} & \textbf{36.35\%} & \textbf{63.65\%} & \textbf{89.08\%} & \textbf{95.00\%} & \textbf{63.65\%} & \textbf{35.23\%} & \textbf{23.60\%} \\
\textbf{GCN-3} & 73.90\% & 37.29\% & 62.71\% & 88.85\% & 94.76\% & 62.71\% & 35.08\% & 23.50\% \\
\textbf{GCN-4} & 73.53\% & 38.36\% & 61.64\% & 87.72\% & 94.05\% & 61.64\% & 34.93\% & 23.58\% \\
\textbf{GCN-5} & 73.41\% & 38.69\% & 61.31\% & 87.50\% & 94.28\% & 61.31\% & 34.84\% & 23.59\% \\
\bottomrule
\end{tabular}
}
\end{table}

The performance of a GCN will differ depending on the number of GCN layers. In order to study how the number of layers affect the performance, we conducted some additional experiments with $1,\ldots,5$ GCN layers, represented by GCN-1, $\ldots$, GCN-5, respectively. Table~\ref{tab:gcn_layer_test} shows the results, which show that the model with two GCN layers achieved the best classification result. In comparison, the model with only one GCN layer showed the worse performance, which can be explained by the too shallow GCN being unable to extract label-to-label correlation effectively. The model's performance dropped while the number of GCN layers increases after two. This is likely caused by overfitting since a too deep GCN may learn about label-to-label correlation too specifically, therefore harming its generalizability. Based on the results, we recommend using two GCN layers for our model.

\subsection{Case Study}

\begin{table}[!htb]
\centering
\caption{Case Study}
\label{tab:case study}
\begin{tabular}{p{0.6\linewidth}c}
\toprule
\makecell[c]{Testing Tweets} & Privacy Aspects\\
\midrule
I have been honored to serve the students in the district that I graduated from for 12 years as a teacher and administrator.
& Occupation, Graduation\\
\midrule
Guys up to now my girlfriend is still looking for my birthday gift and its now two months since then. & Gender\\
\midrule
My trip to Washington D.C. will start in three days. & Place planning to go\\
\midrule
I am only 18 years old and I won’t have enough to cover the removal surgery. & Age, Health condition\\
\midrule
Because I'm a Muslim so there's no way they would let me go to any concert. & Religion\\
\midrule
I have my bachelor's degree in computer science from Stanford university. & Education background\\
\midrule
Choosing who to be included in this project is too stressful. & General complaint\\
\midrule
The Warriors seem to be winning this year's championship. & Neutral statement\\
\bottomrule
\end{tabular}
\end{table}

To demonstrate the practical usefulness of our proposed model, we use several example tweets (not included in the dataset) to demonstrate the effect of the model. To avoid potential privacy disclosures by us, we only use anonymous tweets for this part. For better illustration, the tweets tested try to cover multiple common privacy categories. For clarity, we only present the tweets that are correctly classified by our models.

As Table~\ref{tab:case study} shows, we use several tweets to show the effect of our proposed model, including ten kinds of privacy aspects. For the first seven tweets, our model correctly captured the aspects of the privacy disclosure, which demonstrates the practicality of our proposed model. For example, the third tweet may disclose the travel destination of the user, thus the model outputs ``Place planning to go" as a reminder. The sixth tweet explains where the user obtained their bachelor's degree, so it may disclose the privacy category of ``Education background" according to our model. Therefore, Twitter users and the platform (Twitter) can use these kinds of reminders as a reference to avoid unintended privacy disclosures. For the last tweet, the tweet does not reveal any personal privacy aspect. Therefore, the tweet is classified into the category of ``Neutral statement" by our model.

Furthermore, as Table~\ref{tab:case study} shows, if a tweet may disclose multiple categories of privacy information, our fine-grained privacy disclosure detection model can solve this problem with the consideration of multi-label classification, which shows the advantage of our model compared to other binary coarse-grained privacy disclosure detection models. For example, the detection results of the first testing tweet in Table~\ref{tab:case study} include two privacy aspects: ``Occupation" and ``Graduation", meanwhile the detection results of the fourth tweet include ``Age" and ``Health condition".

\section{Conclusions and Future Work}
\label{sec:conclusion}

A new privacy disclosure detection model is proposed in this paper. The proposed model integrates the text information, the label-to-text correlation and the label-to-label correlation for detecting privacy disclosures in the input text. For the first time, a GCN-assisted feature fusion mechanism is proposed to achieve the text feature fusion process with the guidance of the label-to-label correlation. During the process of feature fusion, the compensation coefficients are proposed to help fuse self-attention and label-attention features. Based on a dataset of privacy-disclosing tweets, our experimental results showed that our model outperformed a number of selected state-of-the-art models and that the improved performance comes from the new design elements we introduced. A number of example tweets are used to demonstrate the practical usefulness of the proposed model. The results show that our proposed model can be used to support development of privacy protection tools that alert online users and online platforms about unintended privacy disclosures.

In our paper, our experiment are based on a single dataset covering 32 privacy-oriented personal aspects~\citep{song2018personal}, considering that this dataset is the best privacy-disclosing dataset we could find. However, using only one single dataset can make it difficult to judge how generalizable our results are. In addition, although the dataset we used covers a rich set of personal aspects, the coverage can still be extended to cover more personal aspects. Therefore, constructing more datasets for privacy disclosure detection is needed so our work can be further validated on multiple datasets. Meanwhile, our model aims to detect the privacy disclosure in text-only UGC. However, non-textual information in UGC such as images and videos can often disclose privacy information, too. Thus, in our future work, we will investigate the construction of a multi-modal privacy disclosure detection model supporting both visual and textual information.

\section{Statements and Declarations}
\label{sec:statements}

\subsection{Funding}

Shujun Li's work was partly funded by the research project ``PRIvacy-aware personal data management and Value Enhancement for Leisure Travellers'' (PriVELT, \url{https://privelt.ac.uk/}), funded by the EPSRC (Engineering and Physical Sciences Research Council), part of the UKRI (UK Research and Innovation), under the grant number EP/R033749/1. Also this work was partly funded by the National Natural Science Foundation of China under the reference number 61972249.

\subsection{Competing Interests}

The authors have no financial or proprietary interests in any material discussed in this article.

\bibliography{main}

\end{document}